**Electron-loss and target ionization cross sections for water vapor by 20-150 keV neutral atomic hydrogen impact**

F. Gobet [a], S. Eden [a], B. Coupier [a], J. Tabet [a], B. Farizon [a], M. Farizon[*][a], M.J. Gaillard [a], S. Ouaskit[†][a], M. Carré [b], T.D. Märk [c]

[a] Institut de Physique Nucléaire de Lyon, IN2P3-CNRS et Université Claude Bernard, 43, Boulevard du 11 novembre 1918, F-69622 Villeurbanne Cedex, France

[b] Laboratoire de Spectrométrie Ionique et Moléculaire, CNRS UMR 5579 et Université Claude Bernard, 43, Boulevard du 11 novembre 1918, F-69622 Villeurbanne Cedex, France

[c] Institut für Ionenphysik, Leopold Franzens Universität, Technikerstr.25
A-6020 Innsbruck, Austria

**Abstract**

A complete set of cross sections is reported for the ionization of water molecules by neutral atomic hydrogen impact at velocities of the order of the Bragg peak. The measured relative cross sections are normalized by comparison with proton impact results for the same target conditions and previous absolute data. Event by event coincidence analysis of the product ions and the projectile enables partial cross sections for target ionization and target plus projectile ionization to be determined, as well as total cross sections for electron loss reactions.

[*] Corresponding author. fax: +33 4 72 44 80 04. email: m.farizon@ipnl.in2p3.fr
[†] Permanent address: Laboratoire de la Matière Condensée, Faculté Ben M'sik, Université Hassan II, Casablanca, Morocco

It is widely accepted today that an understanding of the detailed structure of charged particle tracks is essential to elucidate and interpret the mechanistic consequences of energy deposition by ionizing radiation in an absorbing medium [1]. Concomitant and subsequent radiation damage in the medium (biological tissue) depends crucially upon the evolution of track patterns along the path of the energetic particle. The determination of these patterns relies on a broad range of cross section data for the primary interactions of the relevant particles with the molecular constituents of the absorbing medium [1-4].

A charged particle passing through an absorbing medium at high velocity loses kinetic energy primarily in collisions with bound electrons. Subsequent energy loss by ejected (secondary) electrons and chemical reactions involving energized atoms and molecules produce the track structure and radiation damage [5]. As the charged particle slows down, electron capture interactions become more important thereby producing fast neutral projectiles. The interplay between these different processes, their energy dependences, and the possibility of switching the charge state of the projectile is responsible for the occurrence of the Bragg peak in irradiated media [6]. To develop track structure models, both electron capture and electron loss cross sections are required for each potential charge state of the moving particle. Whereas these cross sections are available for proton impact ([7-13] for water), there are substantial shortcomings in the understanding of the underlying collision processes for neutral hydrogen atom impact, in particular upon biologically relevant molecular targets.

The ionization of water molecules represents both a fundamental example in collision physics [14] and a subject of great interest in various areas of applied physics including radiation damage to biological tissue [15], the chemistry of planetary upper atmospheres [16], and the ionization balance in the plasma of tokamak fusion devices [17]. Today it is recognized that radiation damage to living material, in particular strand breaking in DNA, can occur not only as a result of direct particle-biomolecule interactions but also through processes initiated by radical species released by the fragmentation of neighboring water molecules. A detailed knowledge of the relevant ionization processes is thus essential to achieve a full understanding of radiation damage to biological material on a microscopic level [18]. Information on the individual reactions is required: determination of average energy loss is not sufficient.

In spite of this broad range of interest, cross section measurements for the ionization of water by neutral impact are extremely scarce and a number of decisive details have not been investigated. In 1968 Toburen et al. [8] reported the first *total electron loss cross sections* for 100-2500 keV neutral atomic hydrogen impact on $H_2O$ using the equilibrium fraction method to analyze the multiple-charge-state beam at the exit of a gas cell. Dagnac et al. [19] extended these studies to energies between 2 and 60 keV. In 1986 Bolorizadeh and Rudd [20] measured doubly differential ionization cross sections for electron production in collisions between neutral hydrogen atoms and water molecules at 20-150 keV. The doubly differential cross sections were integrated over angle and energy to obtain *total electron production cross sections*. As the previous experiments did not detect the ionized target system, important information on the ionizing interaction of neutral hydrogen atoms with water vapor is unavailable in the literature.

The present work features the application of an experimental system which enables atomic hydrogen impact ionization of water vapor to be analyzed in great detail and on an event by event basis. Target ion signals can be detected in coincidence with the charge-analyzed projectile to determine whether the ionizing collision of the hydrogen atom involves charge transfer (electron loss) or not (no change in projectile charge state). The present study is possible due to the addition of a time of flight ion detector operated in coincidence with the existing high-energy ion beam / multi-coincidence apparatus [7,21,22]. The fate of the projectile can be analyzed in coincidence with that of the target molecule in each ionizing collision. Thus, for the first time, it is possible to obtain not only *total* ionization cross sections for neutral particle impact but also cross sections that are '*differential*' in terms of the ionization event (target ionization and/or projectile ionization). $H_2O$ is selected as the target due to its key role in radiation damage to biological tissue.

The apparatus has been described previously [7]. However, in order to carry out neutral impact experiments, an argon-filled charge transfer cell is introduced between the magnetic mass selector and the collision cell. Protons are thus converted into neutral hydrogen atoms. Deflection plates situated in front of the collision cell serve to eliminate any residual charged projectiles. This electric field also acts as a quenching field for metastable atoms and an ionizing field for Rydberg atoms [20]. For each single collision event, the product ions produced in the target region are measured in coincidence with the charge of the projectile after the ionizing collision. The detected projectile may be a neutral H atom (target ionization), a proton (target plus projectile ionization), or an $H^-$ ion (electron capture). The latter process is found to be negligible in the present study. Each neutral impact result is accompanied by a proton impact measurement with identical target conditions. Relative cross sections derived from the measured coincident rates are thus calibrated using recently measured relative and absolute proton impact results [7-13]. The maximum errors are estimated at 25% on the basis of the reported errors in the absolute data used for normalization [11] and the deviation of the measured relative cross sections from smooth fitted relations. The increased statistical errors associated with low neutral impact cross section events at low impact energies are diminished by correspondingly high normalizing proton impact cross sections [7]. Naturally, cross sections deduced by subtracting directly measured cross sections are subject to greater percentage errors.

Charge transfer cross sections are generally designated as $\sigma_{if}$ where $i$ is the initial and $f$ the final charge state of the projectile. For example, $\sigma_{01}$ corresponds to the total electron loss cross section for neutral hydrogen impact. However, a more refined nomenclature is required to incorporate the fate of the target. The present partial cross sections are designated as $\sigma_{ii \to ff}$ where the first and second $i$ correspond to the initial charge state of the projectile and the target, respectively. The first and second $f$ stand for the respective final charge states of the projectile and the target. Thus the individual processes studied in the present work are represented as:

$$H + H_2O \to \sigma_{00 \to 01} \to H + \text{(the ionized target system)} \quad (1)$$

$$H + H_2O \to \sigma_{00 \to 11} \to H^+ + e^- + \text{(the ionized target system)} \quad (2)$$

$$H + H_2O \to \sigma_{00 \to 10} \to H^+ + e^- + H_2O \quad (3)$$

Reaction (1) is called *target ionization*, reaction (2) *projectile plus target ionization* or *electron loss plus target ionization*, and reaction (3) *electron loss without target ionization*. That the observed electron loss events are due to interactions with water molecules is confirmed by the negligible detection rate of proton projectiles when the neutral hydrogen beam is tested without the water jet. The present experiment enables the cross sections for reactions (1) and (2) to be measured directly by coincident ion-projectile detection. In addition, the cross section for reaction (3) can be deduced using the total electron loss cross sections $\sigma_{01}$ (non-coincidence measurements carried out previously [8,19] and also here) and the equation:

$$\sigma_{01} = \sigma_{00 \to 11} + \sigma_{00 \to 10} \quad (4)$$

Fig.1 shows the measured total electron loss cross sections $\sigma_{01}$, the measured target ionization cross sections $\sigma_{00 \to 11}$, and the cross sections for electron loss without target ionization $\sigma_{00 \to 10}$ obtained by subtracting $\sigma_{00 \to 11}$ from $\sigma_{01}$. Only the present total electron loss cross sections $\sigma_{01}$ (filled squares) can be compared with earlier determinations, i.e. the data of Dagnac et al. [19] (open squares) and that of Toburen et al. [8] (crossed squares). The present and previous measurements are good agreement, lying comfortably within the estimated error limits. The only exception to this agreement is the 100 eV point from [8] which appears to be systematically shifted to a lower cross section compared to the present study and that of [19].

Moreover, it is now possible to gain a deeper insight into the ionization process by looking at the shape of the individual partial cross section curves $\sigma_{00 \to 11}$ and $\sigma_{00 \to 10}$. Fig.1 shows that the total cross section $\sigma_{01}$ is dominated by reaction (3) at low energies, while reactions (2) and (3) have about the same probability at 50 - 70 keV. At higher energies, reaction (3) is again the major reaction channel.

Besides reactions involving charge transfer, it has also been possible to measure the cross section $\sigma_{00 \to 01}$ for reaction (1): target ionization without a change in the charge state of the projectile. This data is shown in Fig.2 and cannot be compared directly with any previous results. However, the present total cross sections (filled circles in Fig.2) can be compared with the work of Bolorazideh and Rudd [20] (open circles). The total cross section called $\sigma_+$ in the present work (corresponding only to the electrons produced by the ionization of the target), was obtained previously by subtracting the ELC (electron loss to the continuum) from the total electron production cross section ($\sigma_-$) and described as the *total cross section for ionization of the target* [2]. Bolorazideh and Rudd [20] determined the total electron production cross section curve by integration of their doubly differential electron ejection cross sections. The ELC was obtained by the same authors by taking into account that the electrons released from the projectile appear preferentially in the forward direction and essentially with the velocity of the ionized projectile. Accordingly,

$$\sigma_+ = \sigma_- - \text{ELC} = \sigma_- - (\sigma_{00\to10} + \sigma_{00\to11}) \qquad (5)$$

As the total electron production cross section $\sigma_-$ can be written in terms of the individual cross sections as shown below,

$$\sigma_- = \sigma_{00\to01} + 2\sigma_{00\to11} + \sigma_{00\to10} \qquad (6)$$

by inserting (6) into (5) for $\sigma_+$ the following simple relationship is obtained

$$\sigma_+ = \sigma_{00\to01} + \sigma_{00\to11} \qquad (7)$$

Fig.2 shows $\sigma_{00\to01}$, $\sigma_{00\to11}$, and $\sigma_+$. The errors on the previous values for $\sigma_+$ are given as 20% [20]. Therefore, the measured $\sigma_+$ agree with the previous work to within the estimated error limits except at the low and high extremes of the energy range of both sets of experiments.

The present absolute values can also be compared with the total electron production cross sections $\sigma_-$ reported by Bolorazideh and Rudd [20] and shown as open triangles in Fig.3. Using (5) we can rewrite equation (6)

$$\sigma_- = \sigma_{00\to 01} + \sigma_{00\to 11} + \sigma_{00\to 11} + \sigma_{00\to 10} = \sigma_+ + \sigma_{01} \qquad (8)$$

thereby expressing $\sigma_-$ in terms of the presently determined partial cross sections. Thus it is unsurprising that, as observed for $\sigma_+$, Fig.3 shows the presently derived values for $\sigma_-$ and those measured by Bolorizadeh and Rudd [20] to lie within the estimated error limits except at the low and high extremes of the energy range.

Finally, it is instructive to compare the present neutral hydrogen impact results with earlier data [7-13] for proton impact ionization of water. The cross sections for direct target ionization (without electron capture) by proton impact $\sigma_{10\to 11}$ and for total target ionization by neutral impact $\sigma_+$ are approximately equal (maximum of $5\times 10^{-16}$ cm$^2$ at 65 and 80 keV, respectively). Therefore, the high total target ionization cross sections at lower energies for proton impact compared to neutral impact are due to the contribution of electron capture by the projectile. In contrast, the cross section for charge-state change (electron loss) by the neutral projectile $\sigma_{01}$ decreases with impact energy below 95 keV. The dependence of the cross sections for ionization processes upon changes in the projectile charge state is a crucial determinant for the track structure along the path of the particle and thus significantly affects the damage to the absorbing medium [2].

The present work provides the first results recorded on an event by event basis for atomic hydrogen impact ionization of water vapor. In addition to the determination of average cross sections, experiments of this kind can explicit the fluctuations of individual processes, thus opening the possibility for theoretical treatments of hydrogen impact ionization in a medium which go beyond the presently available analytical functions [2,23]. In the present analysis the results provide, to the authors' knowledge, the first complete set of cross sections for the ionization of a molecular target by neutral hydrogen impact including various total and partial cross sections, and, furthermore, differentiating between the direct ionization and the electron loss mechanisms.

**Acknowledgements**

This research was supported by the FWF, ÖNB and ÖAW, Vienna, the European Commission, Brussels, and the French and Austrian governments through the Amadee and PICS programs.

Figure captions

Fig.1

Events involving electron loss by the projectile ($H^+$ detection) in neutral atomic hydrogen impact upon water vapor: total electron loss cross section $\sigma_{01}$ and projectile plus target ionization cross section $\sigma_{00\to 11}$. The cross section for electron loss without target ionization $\sigma_{00\to 10}$ is deduced by subtracting $\sigma_{00\to 11}$ from $\sigma_{01}$.

Fig.2

Events involving target ionization in neutral atomic hydrogen impact upon water vapor: cross sections for projectile plus target ionization $\sigma_{00\to 11}$, target ionization only (with $H^0$ detection) $\sigma_{00\to 01}$, and the total cross section for target ionization $\sigma_+$ (equal to $\sigma_{00\to 11} + \sigma_{00\to 01}$).

Fig.3

Events involving electron emission in neutral atomic hydrogen impact upon water vapor: total cross sections for target ionization $\sigma_+$ and for electron loss by the projectile $\sigma_{01}$. The total electron emission cross section $\sigma_-$ is deduced by summing $\sigma_{01}$ and $\sigma_+$.

Fig.1

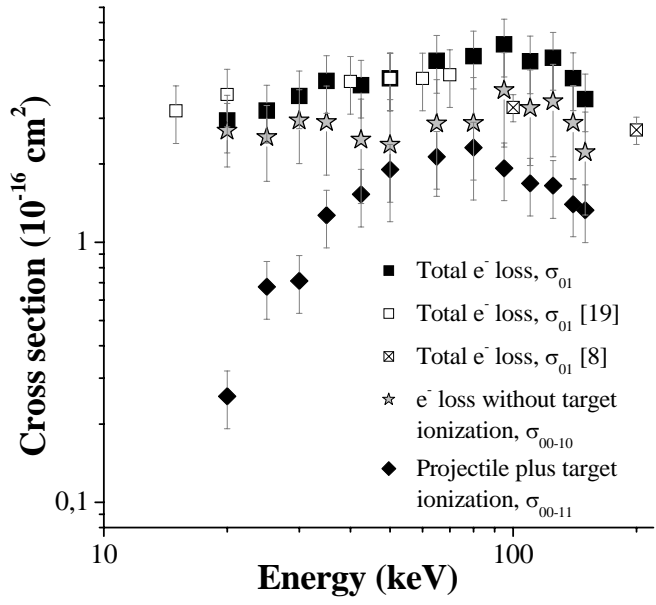

- ■ Total e⁻ loss, $\sigma_{01}$
- □ Total e⁻ loss, $\sigma_{01}$ [19]
- ⊠ Total e⁻ loss, $\sigma_{01}$ [8]
- ☆ e⁻ loss without target ionization, $\sigma_{00-10}$
- ◆ Projectile plus target ionization, $\sigma_{00-11}$

Fig.2

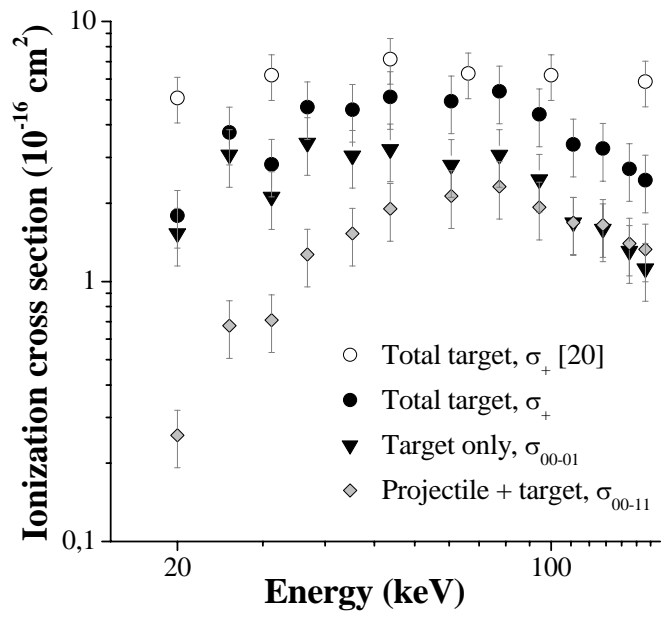

Fig.3

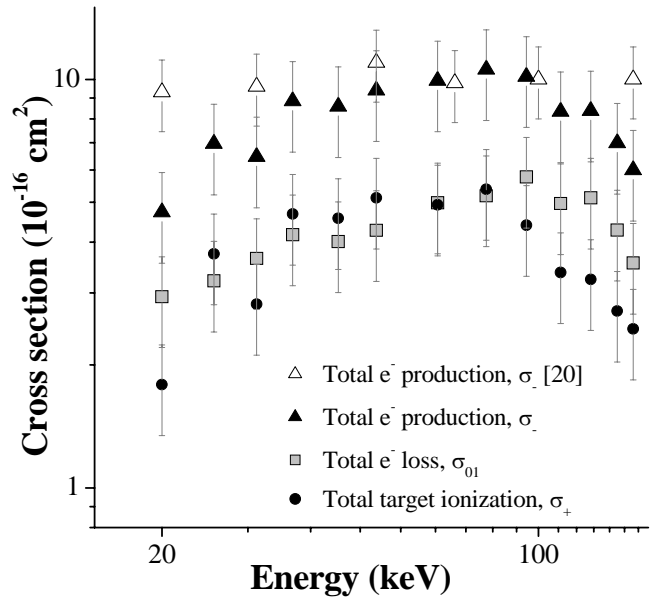

△ Total e⁻ production, $\sigma_-$ [20]
▲ Total e⁻ production, $\sigma_-$
▪ Total e⁻ loss, $\sigma_{01}$
● Total target ionization, $\sigma_+$